\documentclass[final, authoryear,3p,12pt]{elsarticle}

\usepackage{amssymb}
\usepackage{amsthm}

\usepackage{mathrsfs}
\usepackage{float}
\usepackage{eufrak}

\usepackage{graphics}
\usepackage{amsfonts}
\usepackage{mathrsfs}
\usepackage{amscd}
\usepackage{color}
\usepackage{url}
\usepackage[plainpages=false,pdfpagelabels=true,colorlinks=true,citecolor=blue,hypertexnames=false]{hyperref}
\usepackage[ruled,vlined,boxed,commentsnumbered]{algorithm2e}

\journal{}

\newtheorem{theorem}{Theorem}[section]

\newtheorem{remark}[theorem]{Remark}

\def\ker{{\rm Kernel}}
\def\image{{\rm Image}}
\def\span{{\rm Span}}
\def\syz{{\rm Syzygy}}

\def\lm{{\rm lm}} 
\def\lc{{\rm lc}}

\def\e{{\bf e}}
\def\m{{\bf m}}

\def\u{{\bf u}}
\def\v{{\bf v}}
\def\w{{\bf w}}

\def\lcm{{\rm lcm}}

\def\max{{\rm max}}

\def\lla{{\longleftarrow}}

\newcommand{\comment}[1]{}
\newcommand{\ignore}[1]{}

\begin{document}

\begin{frontmatter}



\title{Signature-Based Gr\"obner Basis Algorithms --- Extended MMM Algorithm for computing Gr\"obner bases}


\author{Yao Sun}
\ead{sunyao@iie.ac.cn}




\address{SKLOIS, Institute of Information Engineering, CAS, Beijing 100093,  China}


\begin{abstract}
Signature-based algorithms is a popular kind of algorithms for computing Gr\"obner bases, and many related papers have been published recently. In this paper, no new signature-based algorithms and no new proofs are presented. Instead, a view of signature-based algorithms is given, that is, signature-based algorithms can be regarded as an extended version of the famous MMM algorithm. By this view, this paper aims to give an easier way to understand signature-based Gr\"obner basis algorithms. 
\end{abstract}


\end{frontmatter}



\section{Introduction} \label{sec_intro}

Gr\"obner basis has been shown to be a powerful tool of solving systems of polynomial equations as well as many important problems in algebra.

\subsection{Improvements of Gr\"obner basis algorithms}

Since Gr\"obner basis is proposed in 1965 \citep{Buch65}, many improvements have been made to speed up algorithms for computing Gr\"obner bases. These improvements can be concluded into the following three kinds.

\begin{enumerate}

\item {\bf Detecting redundant computations/critical pairs.}

During the computation of a Gr\"obner basis, redundant computations usually refer to  computations of reducing polynomials to 0, because this kind of computations makes no contribution to build a Gr\"obner basis (in signature-based algorithms, reducing a polynomial to 0 may contribute to build a Gr\"obner basis for the syzygy module). 

The first criteria for detecting redundant computations are proposed by Buchberger \citep{Buchberger79}. Syzygies of polynomials are first used to detect useless computations in \citep{moller92}. Faug\`ere proposes an improved version of syzygy criterion by using principal syzygies in his famous F5 algorithm \citep{Fau02}, and claims almost all redundant computations are rejected. Criteria presented in \citep{Gao10b} as well as \citep{Arri11} can detect a bit more redundant computations, since besides using the information of principal syzygies, they also use non-principal syzygies obtained during the computation of Gr\"obner bases.

\item {\bf Speeding up necessary computations.}

The most fundamental operation in computing a Gr\"obner basis is polynomial reduction, or more specifically, polynomial additions and monomials times polynomials. Faug\`ere has said, during the computation of a Gr\"obner basis, almost all time are spent on reducing polynomials. Thus, speeding up the efficiency of basic polynomial operations will improve the whole algorithm significantly. 

Linear algebraic techniques are introduced to do polynomial reductions after Lazard points out the relation between a Gr\"obner basis and a linear basis of an ideal \citep{Lazard83}. Gebauer-M\"oller algorithm can be regarded as an implementation of Lazard's idea \citep{Gebauer86}. Lazard's idea also leads to the famous F4 algorithm \citep{Fau99} and XL algorithm \citep{Courtois00}. In boolean polynomial ring, zdd (zero-suppressed binary decision diagram) is introduced to optimize the basic operations of boolean polynomials \citep{Brick09}.

\item {\bf Finding appropriate parameters/strategies.}

It is known that monomial orderings used in a Gr\"obner basis algrotihm affects the efficiency a lot. Now, it is commonly believed that the graded reverse lexicographic orderings usually has the best performance for computing a Gr\"obner basis. 

The strategies for choosing critical pairs/S-polynomials also play important roles in a Gr\"obner basis algorithm, because these strategies decide which polynomials are reduced before others. Buchberger's third criterion \citep{Buchberger79} suggest reducing critical pairs/S-polynomials with the smallest degree first. This criterion seems to be most efficient strategy in many examples, so it is now used in most Gr\"obner basis algorithms, including F5. Giovini et al.'s algorithm chooses critical pairs/S-polynomial by ``sugar'' \citep{Gio91}. Some signature-based Gr\"obner basis algorithms choose critical pairs or J-pairs (equivalent to critical pairs) with the smallest signature.  

In algorithms dealing with critical pairs in a batch, for example F4 and F5, how many critical pairs are handled at a time is also a question. Faug\`ere suggests dealing with all the critical pairs with the smallest degree at a time.

\end{enumerate}

\subsection{Signature-based Gr\"obner basis algorithms}

F5, proposed by Faug\`ere, is the first signature-based Gr\"obner basis algorithm \citep{Fau02}. F5 is considered as the most efficient algorithm at present, and F5 has even successfully attacked many famous cryptosystems, including HEF \citep{Fau03}.

Original F5 is written in pseudo-codes, and its proofs, such as the correctness and termination, are not given completely. So F5 seems very complicated to understand for a long time. There are few papers studying the theoretical aspects of F5 before the year 2008, except Stegers' thesis \citep{Stegers05}, in which Stegers rewrites F5 in more detail, but no new proofs are included.

Eder's paper \citep{Eder08} may be the first paper studying the correctness of F5, and is available online in 2008. Motivated by Eder's ideas, the authors begin to study F5 in a more general sense. Orginal F5 assumes the input polynomials are homogeneous, and it is also written in an incremental style, i.e., firstly computing a Gr\"obner basis for $\langle f_1\rangle$, then secondly a Gr\"obner basis for $\langle f_1, f_2\rangle$, $\cdots$,  and finally a Gr\"obner basis for $\langle f_1, \cdots, f_{m}\rangle$. However, the authors notice F5 in this fashion cannot work efficiently for cryptosystems. That is, polynomials in boolean rings are not homogeneous, and in many examples, such as the HFE cryptosystem, a Gr\"obner basis for $\langle f_1\rangle$ over a boolean polynomial ring is very expensive to compute than a Gr\"obner basis for $\langle f_1, \cdots, f_{m}\rangle$. Besides, if F5 works incrementally, the inputing order of polynomials $f_1, f_2, \cdots, f_m$ affects the efficiency significantly. On seeing this, the authors start to change original F5 to another fashion. Firstly, the authors rewrite F5 equivalently in a style similar to Buchberger's classical algorithm. In this algorithm (called F5b), original F5 can be obtained easily by choosing some parameters in F5b.  Moreover, inputing polynomials are not required to be homogeneous, and this algorithm can also work non-incrementally. Secondly, the authors prove the correctness of F5b, and finally propose a variant of F5 which has fewer dependence on the ordering of inputing polynomials. These result are first published in \citep{SunWang09a} and \citep{SunWang09c}, and then reported in \citep{SunWang09b}. A polished version is available online in \citep{SunWang10}, and finally published in \citep{SunWang11a} and \citep{SunWang13a}.

Later, from private communications with Professor Faug\`ere, the authors learn that original F5 requiring homogeneous inputs and written in an incremental fashion is just for simplicity. F5 can work both incrementally and non-incrementally since it is proposed, and F5 also computes critical pairs with the smallest degree even for non-homogeneous inputs. 

In the year 2009, another two important variants of F5, called F5c and F5e respectively, are also proposed independently on MEGA 2009, and the versions with detailed proofs are published in the special issue of MEGA \citep{Eder10} and \citep{Ars10}. In the algorithm F5c, Eder and Perry optimize the incremental version of F5. Specifically, F5c uses the reduced Gr\"obner basis of $\langle f_1, \cdots, f_{m-1}\rangle$ to compute a Gr\"obner basis $\langle f_1, \cdots, f_{m}\rangle$, which will avoid many redundant computations. Eder and Perry also give a complete proof for the correctness of F5c, and their implementation of F5c is regarded as standard comparisons of following papers. The idea of Hashemi and Ars' F5e is quite similar to the authors' variant F5 algorithm  proposed in \citep{SunWang09c}.  F5e aims to make F5 have fewer influence on the computing order of inputing polynomials, and hence, can work non-incrementally. However, Gao et al. point out in \citep{Gao10b} that proofs published in \citep{Ars10} have minor errors.

In 2010, Gao et al. report their G2V algorithm on ISSAC 2010 \citep{Gao10a}. G2V is also an incremental algorithm for computing Gr\"obner bases. The feature of G2V is that, it can compute Gr\"obner bases for both $\langle f_1, f_2, \cdots, f_{m-1}\rangle: f_m$ and $\langle f_1, f_2, \cdots, f_m\rangle$ at the same time when a Gr\"obner basis for $\langle f_1, f_2, \cdots, f_{m-1}\rangle$ is known. No proofs for this algorithm is presented in that paper, but timings are very catching, which seems much faster than timings reported in \citep{Eder10}. Later in 2010, Gao et al. put their GVW algorithm online \citep{Gao10b}. GVW is also a signature-based Gr\"obner basis algorithm, and gives a different view of all signature-based algorithms. We will present detailed discussions on GVW in current paper sooner.

Since F5 and GVW are both signature-based Gr\"obner basis algorithms, researchers begin to study the similarity between F5 and GVW in order to reveal the essence of signature-based algorithms. Huang put his paper online in November of 2010 \citep{Huang10}.  In his paper, Huang proposes a new structure of signature-based algorithms, and shows which kind of polynomials have to be computed. Moreover, Huang also gives a method of proving the termination of signature based algorithms, and termination of original GVW is also proved. On the other side, the authors generalize criteria in F5 and GVW, and show which kind of redundant computations can be rejected correctly in signature-based algorithms \citep{SunWang11b}. Eder-Perry gives a new structure to ensure signature-based algorithms terminate in finite steps \citep{Eder11b}, which is an extension of their previous work \citep{Eder11a}.

On criteria of GVW, after noticing original GVW's ``eventually super reducible criterion'' is not efficient. An improved criterion is proposed independently almost at the same time \citep{Huang10, SunWang11b, Arri11}.

In 2011, there is almost no doubts about the correctness of signature-based algorithms. Researchers turn to study the termination. Early proofs on termination assume critical pairs or JPairs (in GVW) are handled by an incremental order on signatures.\footnote{In some papers, ordering on signatures is assumed to be ``degree compatible'' ordering, and critical pairs with smallest degrees are dealt with first. It is easy to prove this assumption is equivalent to assuming ``critical pairs are handled by an incremental order on signatures''. }  Termination of GVW is first proved with this assumption in \citep{Huang10}, and later proved without this assumption in \citep{SunWang12}. Termination of original F5 is still unproved now. Since in original F5, a polynomial is rewritten only by the polynomial generated later than it, this ``generating order'' condition is hardly used in the proof of termination because it gives few information on monomials. The termination of variants of F5 have been studied in \citep{Eder11a, Eder11b, Arri11, Galkin12a, Galkin12b, Pan13}.

Regarding to implementations of signature-based algorithms, Faug\`ere's F5 implementation have been proven to be the most efficient implementation, and it also has a parallel version \citep{Fau10}. Roune et al.'s implementation of GVW and Arri-Perry algorithm is also very efficient \citep{Roune12}.

There still many other related works on signature-based algorithms. Zobnin discusses F5 in a matrix form \citep{Zobnin10}. Sun and Wang extend signature-based algorithms to compute Gr\"obner bases for differential operators \citep{SunWang12}, solve detachability problems in polynomial rings \citep{SunWang11c}, and extend GVW to compute more Gr\"obner bases \citep{SunWang13b}. Eder extends  signature-based algorithms to compute standard bases \citep{Eder12a}, analyzes inhomogeneous Gr\"obner basis computations \citep{Eder12b}, and improves incremental algorithms \citep{Eder13}.  Gertdt and Hashemi apply Buchberger's criteria to signature-based algorithms \citep{Gerdt13}.

\subsection{Contributions in current paper}
 
The authors are not going to give new algorithms or new proofs on signature-based Gr\"obner basis algorithm. Instead, the authors try to present a simpler view of GVW as well as all signature-based algorithms, hoping to make signature-based algorithms easier understood. We guess some existing signature-based algorithms are developed in the same way as described in this paper, but in order to be more precise and rigorous, these algorithms are not presented in this way. This paper will mainly talk about the ideas how signature-based algorithms are developed, and may not be so rigorous in mathematics in some places.

The authors will introduce MMM algorithm first \citep{mmm92}, which can be regarded as a generalized algorithm of FGLM \citep{fglm93}.  Then we will show how to deduce the GVW algorithm from MMM. This paper is organized as follows. The MMM algorithm and related notations are introduced in Section\ref{sec_mmm}. We show how GVW is deduced from MMM in Section \ref{sec_gvw}. Concluding remarks follow in Section \ref{sec_conclusions}.

\section{The MMM algorithm} \label{sec_mmm}

Let $k[X] := k[x_1, \ldots, x_n]$ be a polynomial ring over a field $k$ with $n$ variables $X = \{x_1, \ldots, x_n\}$. Given a monomial order $\prec$ on $k[X]$, for a polynomial $f=c_1x^{\alpha_1} + \cdots + c_tx^{\alpha_t} \in k[X]$ where $c_i \in k$ and $i=1, \ldots, t$, the leading monomial and leading coefficient of $f$ w.r.t. $\prec$ is defined as $\lm(f) := x^{\alpha_k}$ and $\lc(f) := c_k$, where $x^{\alpha_k} = \max_{\prec}\{x^{\alpha_i} \mid c_i \not=0, i=1, \ldots, t\}$.

\subsection{Basic ideas}

The FGLM algorithm is a very efficient algorithm for changing Gr\"obner basis monomial orderings in 0-dimensional ideals. The MMM algorithm generalizes the FGLM algorithm to compute more Gr\"obner bases by using a $k$-linear map $$L: k[X] \longrightarrow V,$$  where $V$ is a $k$-vector space with finite dimension. The MMM algorithm will compute a Gr\"obner basis for the ideal $$\ker(L) = \{f \in k[X] \mid L(f) = 0\},$$ for any given monomial ordering.

In fact, MMM algorithm uses {\em an enumerating method} to find all monomials in $$\lm(\ker(L)) = \{\lm(f)\mid f\in \ker(L)\},$$ as well as all polynomials in a Gr\"obner basis of $\ker(L)$. We can briefly write main ideas of MMM algorithm through the following simple algorithm.

\smallskip\smallskip\smallskip
\noindent{\bf Input:} $L$, a $k$-linear map from $k[X]$ to a finite dimensional vector space $V$; $\prec$, a monomial ordering on $k[X]$.

\noindent{\bf Output:} A Gr\"obner basis of $\ker(L)$ w.r.t. $\prec$.

\begin{enumerate}

\item Sorting all monomials in $k[X]$ by an ascending order on $\prec$: $$m_0 \prec m_1 \prec \cdots \prec m_i \prec \cdots, $$ where $m_i$ is a monomial in $k[X]$.

\item $m_i$'s are proceeded repeatedly according to the above ascending order.

\item For each $m_i$, checking whether $L(m_i)$ is a $k$-linear dependent with $\{L(m_0), L(m_1)$, $\ldots, L(m_{i-1})\}$ in $V$.

\item If $L(m_i)$ is $k$-linear dependent with $\{L(m_0), L(m_1), \ldots, L(m_{i-1})\}$, then there exist $c_0, c_1, \ldots, c_{i-1} \in k$, such that $$L(m_i) = c_0L(m_0) + c_1L(m_1) + \cdots + c_{i-1}L(m_{i-1}),$$ which means $$m_i - (c_0m_0 + \cdots + c_{i-1}m_{i-1}) \in \ker(L) \mbox{ and } m_i \in \lm(\ker(L)),$$ since $L$ is a $k$-linear map.

\item Goto step 2 unless all monomials in $k[X]$ are considered.

\end{enumerate}

Obviously, there is no doubt about the correctness of the above simple algorithm, but there are two problems to be settled.

\begin{enumerate}

\item Generally, there are infinite monomials in $k[X]$, so we cannot enumerate them all. This means the above algorithm does not always terminate.

\item How to check linear dependency at step 3 and compute $c_i$'s at step 4 efficiently?

\end{enumerate}

We show methods of solving the above two problems in the next two subsections  respectively.

\subsection{To ensure termination: syzygy criterion}

If $L(m_i)$ is $k$-linear dependent with $\{L(m_0), L(m_1), \ldots, L(m_{i-1})\}$, i.e.  there exist $c_0, c_1, \ldots, c_{i-1} \in k$, such that $$L(m_i) = c_0L(m_0) + c_1L(m_1) + \cdots + c_{i-1}L(m_{i-1}),$$ then for any $m_k = tm_i$, where $t$ is a monomial in $k[X]$, we have $$L(m_k) = L(tm_i) = c_0L(tm_0) + c_1L(tm_1) + \cdots + c_{i-1}L(tm_{i-1}),$$ which means $$m_k - (c_0tm_0 + \cdots + c_{i_1}tm_{i-1}) \in \ker(L) \mbox{ and } m_k \in \lm(\ker(L)).$$

Since $m_i - (c_0m_0 + \cdots + c_{i-1}m_{i-1})\in \ker(L)$ has been obtained, the polynomial $m_k - (c_0tm_0 + \cdots + c_{i-1}tm_{i-1})$ is no longer needed in a Gr\"obner basis of $\ker(L)$. Thus, we can skip all monomials $tm_i$ in the algorithm when $m_i\in \lm(\ker(L))$. We call this criterion {\em syzygy criterion of MMM}, in order to be consistent with the syzygy criterion of GVW.

\subsection{To check linear dependency: a linear basis of $\image(L)$}

A general way for checking linear dependency is to compute a linear basis. Assume $B_{i-1}$ is a $k$-linear basis of $\span\{L(m_0), \ldots, L(m_{i-1})\}$, which is the vector space generated by $\{L(m_0), \ldots, L(m_{i-1})\}$.
Using the general linear reduction/elimination in $V$, we have the following facts.

\begin{enumerate}

\item If $L(m_i)$ is linear reduced to $0$ by $B_{i-1}$, then we have $m_i \in \lm(\ker(L))$.

\item If $L(m_i)$ is linear reduced to $v\not=0$ by $B_{i-1}$, then $\{v\} \cup B_{i-1}$ is a linear basis of $\span\{L(m_0), \ldots$, $L(m_{i-1}), L(m_i)\}$. 

\end{enumerate}
In the former case,  the multiples of $m_i$ are not considered according to the sysygy criterion of MMM; in the latter case, the dimension of $\span\{L(m_0), \ldots$, $L(m_{i-1}), L(m_i)\}$ is enlarged. This ensures the termination of MMM, since $V$ is a finite vector space. Besides, please note that the linear basis is also updated in the latter case.

In order to obtain the coefficients $c_i's$, preimages of elements in $B_{i-1}$ should also be kept in the algorithm. That is, for each $v \in B_{i-1}$, we should store $u \in k[X]$ such that $L(u) = v$. For such a pair $(u, v)$, $\lm(u)$ is called the signature of this pair.

\begin{remark}
In fact, the complete expression of $u$ does not have to be stored in the algorithm. Instead, we only need to record $\lm(u)$, and the full expression of $u$ can be recovered after the algorithm terminates, by a similar method in signature-based algorithms. This method will be discussed later.
\end{remark}

\section{The GVW algorithm} \label{sec_gvw}

\subsection{From MMM to GVW} \label{subsec_frommmm}

From discussions in last section, we can see that the MMM algorithm actually computes $$\mbox{a Gr\"obner basis for } \ker(L) \mbox{ and a $k$-linear basis for } \image(L),$$ at the same time.

In the GVW algorithm, relations between signatures and corresponding polynomials can be concluded as a homomorphism. Specifically, the following $k[X]$-homomorphism is used in GVW: $$\varphi: k[X]^m \longrightarrow k[X],$$ $$\u = (p_1, p_2, \ldots, p_m) \longmapsto f = p_1f_1 + p_2f_2 + \cdots + p_mf_m,$$ where $f_1, \ldots, f_m\in k[X]$ are given polynomials. The map $\varphi$ is a $k[X]$-homomorphism, since for any $\u, \v\in k[X]^m$ and $p\in k[X]$ we have $$\varphi(\u+\v) = \varphi(\u) + \varphi(\v) \mbox{ and } \varphi(p\u) = p\varphi(\u).$$

The GVW algorithm actually computes $$\mbox{Gr\"obner bases for } \ker(\varphi) \mbox{ and } \image(\varphi),$$ at the same time, where $$\ker(\varphi) = \syz(f_1, \ldots, f_m) = \{\u \in k[X]^m \mid \varphi(\u) = 0\}$$ and $$\image(\varphi) = \langle f_1, \ldots, f_m\rangle.$$

If we generalize this homomorphism $\varphi$, we can extend GVW algorithm to compute more Gr\"obner bases. This work is presented in \citep{SunWang13b}.

\subsection{GVW in MMM style} \label{subsec_gvwinmmm}

First, we write GVW in an MMM style, and deduce the true GVW algorithm afterwards.

\smallskip\smallskip\smallskip
\noindent{\bf Input:} $\varphi$, the $k[X]$-homomorphism from $k[X]^m$ to $k[X]$, defined by $\{f_1, \ldots, f_m\}$ in the last subsection; $\prec_s$ and $\prec_p$, monomial orderings on $k[X]^m$ and $k[X]$ respectively.

\noindent{\bf Output:} Gr\"obner bases of $\ker(\varphi)$ and $\image(\varphi)$ w.r.t. $\prec_s$ and $\prec_p$ respectively.

\begin{enumerate}

\item Sorting all monomials in $k[X]^m$ by an ascending order on $\prec_s$: $$\m_0 \prec_s \m_1 \prec_s \cdots \prec_s \m_i \prec_s \cdots, $$ where $\m_i = x^\alpha\e_j$ is a monomial in $k[X]^m$ and $\e_j$ is the $j$th-unit.

\item $\m_i$'s are proceeded repeatedly according to the above ascending order.

\item For each $\m_i$, checking whether $\varphi(\m_i)$ is a $k$-linear dependent with $\{\varphi(\m_0), \varphi(\m_1), \ldots$, $\varphi(\m_{i-1})\}$ in $k[X]$.

\item If $\varphi(\m_i)$ is $k$-linear dependent with $\{\varphi(\m_0), \varphi(\m_1), \ldots, \varphi(\m_{i-1})\}$, then there exist $c_0, c_1, \ldots, c_{i-1} \in k$, such that $$\varphi(\m_i) = c_0\varphi(\m_0) + c_1\varphi(m_1) + \cdots + c_{i-1}\varphi(\m_{i-1}),$$ which means $$\m_i - (c_0\m_0 + \cdots + c_{i-1}\m_{i-1}) \in \ker(\varphi) \mbox{ and } \m_i \in \lm(\ker(\varphi)),$$ since $\varphi$ is a $k[X]$-homomorphism.

\item Goto step 2 unless all monomials in $k[X]^m$ are considered.

\end{enumerate}

Clearly, it is easy to prove that the above algorithm will correctly compute a Gr\"obner basis for $\ker(\varphi)$ and a $k$-linear basis for $\image(\varphi)$, which is also a Gr\"obner basis of $\image(\varphi)$. But there are still several problems to be settled.

\begin{enumerate}

\item Since there are infinite monomials in $k[X]^m$ generally, it is impossible to enumerate them all.

\item The linear dimension of $\image(\varphi)$ is infinite.

\item Checking linear dependency cost too much time and space when the linear dimension of $\{\varphi(\m_0), \varphi(\m_1), \ldots, \varphi(\m_{i-1})\}$ is huge.

\end{enumerate}

Similarly to what we have done in the last section, we show how these problems are settled in GVW in the following subsections.

\subsection{GVW syzygy criterion}

The syzygy criterion of MMM still works, and it is just the GVW syzygy criterion. That is, if $\varphi(\m_i)$ is $k$-linear dependent with $\{\varphi(\m_0), \varphi(\m_1), \ldots, \varphi(\m_{i-1})\}$, i.e. $\m_i \in \lm(\ker(\varphi))$,  then $\m_k = t\m_i \in \lm(\ker(\varphi))$ for any monomial $t$ in $k[X]$. Thus, all monomials like $t\m_i$ can be skipped in the algorithm.

\subsection{Replacing linear bases by strong Gr\"obner bases}

Let $\image_{i - 1}(\varphi)$ denote the $k$-vector space $\span\{\varphi(\m_0), \ldots, \varphi(\m_{i-1})\}$. Please note that $\image_{i-1}(\varphi)$ also contains all the images of polynomials with smaller leading monomials than $\m_i$ in $k[X]$.

Storing a $k$-linear basis $B_{i-1}$ of $\image_{i -1}(\varphi)$ usually takes too much space. So we prefer to using a smaller subset of $B_{i-1}$, which can also be used for checking whether $\varphi(\m_i)$ is in $\image_{i-1}(\varphi)$. We call a set $G_{i-1}$ a {\bf strong Gr\"obner basis} \footnote{This definition of strong Gr\"obner basis is slightly different from that in \citep{Gao10b}, because elements like $0 = \varphi(\v)$ are not required in this strong Gr\"obner basis.}of $\image_{i-1}(\varphi)$, if 
\begin{enumerate}

\item $G_{i-1} = \{g_1 = \varphi(\v_1), g_2 = \varphi(\v_2), \ldots, g_s = \varphi(\v_s)\}$ is a subset of $\image_{i-1}(\varphi)$, and

\item $\image_{i-1}(\varphi)$ is spanned by $\{tg \mid  g = \varphi(\v) \in G_{i-1}$ and $t$ is a monomial in $k[X]$ such that $\lm(t\v) \prec_s \m_i \}$.

\end{enumerate} 

Clearly, a linear basis of $\image_{i-1}(\varphi)$ is a strong Gr\"obner basis of $\image_{i-1}(\varphi)$, but a strong Gr\"obner basis could contain fewer polynomials than a linear basis.

A strong Gr\"obner basis $G_{i-1}$ of $\image_{i-1}(\varphi)$ can be used to check whether $\varphi(\m_i)$ lies in $\image_{i-1}(\varphi)$, because $G_{i-1}$ has the following property. That is, for any $f\in \image_{i-1}(\varphi)$, there always exists $g = \varphi(\v) \in G_{i-1}$ and a monomial $t \in k[X]$ such that 
\begin{enumerate}

\item $\lm(tg) = \lm(f)$, and 

\item $\lm(t\v) \prec_s \m_i$.

\end{enumerate}
Please note that $\lm(tg)$ is the leading monomial w.r.t. $\prec_p$, and $\lm(t\v)$ is the leading monomial w.r.t. $\prec_s$.

Thus, checking whether $\varphi(\m_i)$ lies in $\image_{i-1}(\varphi)$, we can use the following reduction. For $f\in k[X]$, we say $f$ is {\bf reducible} by $G_{i-1}$, if there exists $g = \varphi(\v) \in G_{i-1}$, such that 
\begin{enumerate}

\item $\lm(g)$ divides $\lm(f)$, and

\item $\lm(t\v) \prec_s \m_i$, where $t = \lm(f)/\lm(g)$.

\end{enumerate}
If $f$ is reducible by such $g=\varphi(\v)$, we say $f \longrightarrow_{G_{i-1}} f - ctg = f - ct\varphi(\v)$ is a one-step-reduction of $f$ by $G_{i-1}$, where $c = \lc(f)/\lc(g)$ and $t = \lm(f)/\lm(g)$. We say $f \longrightarrow_{G_{i-1}} f^*$, if $f^*$ is obtained by successive one-step-reductions from $f$ by $G_{i-1}$, and $f^*$ is not reducible by $G_{i-1}$.

Doing reduction to $\varphi(\m_i)$ by $G_{i-1}$, we will get the following cases.

\begin{enumerate}

\item If $\varphi(\m_i) \longrightarrow_{G_{i-1}} 0$, then by definition, there exist $p_1, \ldots, p_s\in k[X]$ such that $$\varphi(\m_i) = p_1g_1 + \cdots + p_sg_s =  p_1\varphi(\v_1) + \cdots + p_s\varphi(\v_s),$$ where $G_{i-1} = \{g_1 = \varphi(\v_1), \ldots, g_s = \varphi(\v_s)\}$ and $\lm(p_j\v_j)\prec_s \m_i$. This means $$\m_i - (p_1\v_1 + \cdots + p_s\v_s) \in \ker(\varphi) \mbox{ and } \m_i \in \lm(\ker(\varphi)).$$

\item If $\varphi(\m_i) \longrightarrow_{G_{i-1}} h \not=0$, then there are two possible cases depending on whether $h$ plays a role in a strong Gr\"obner basis of $\image_i(\varphi)$.
\begin{enumerate}

\item If there exists $g=\varphi(\v)\in G_{i-1}$ such that \begin{equation}\label{equ_super}
\lm(g) \mid \lm(h) \mbox{ and } \lm(t\v) = \m, \mbox{ where } t = \lm(h)/\lm(g),
\end{equation} then $G_{i-1}$ is a strong Gr\"obner basis of $\image_i(\varphi)$.

\item If there is no such $g=\varphi(\v)\in G_{i-1}$ satisfying conditions in (\ref{equ_super}), then $\{h\} \cup G_{i-1}$ is a strong Gr\"obner basis of $\image_i(\varphi)$.

\end{enumerate}

\end{enumerate}

Thus, by doing reduction to $\varphi(\m_i)$, a strong Gr\"obner basis of $\image_i(\varphi)$ can also be obtained, such that the reduction can be done to $\varphi(\m_{i+1})$ sooner. However, reductions in case (a) is redundant, because it makes no contribution to building either a Gr\"obner basis of $\ker(\varphi)$ or a strong Gr\"obner basis of $\image(\varphi)$. Thus, in GVW, reductions in case (a) are rejected by the ``eventually super reducible'' criterion, which is later improved in \citep{Huang10, SunWang11b, Arri11}.

Note that a strong Gr\"obner basis of $\image_\infty(\varphi) = \image(\varphi)$ is also a Gr\"obner basis of $\image(\varphi)$ w.r.t. $\prec_p$.

\subsection{Reducing a simpler form of $\varphi(\m_i)$} \label{subsec_simpler}

Regarding to $\varphi(\m_i)$, if there exist $\varphi(\v)\in G_{i-1}$ and a monomial $t\in k[X]$, such that $\lm(t\v) = \m_i$, then it is easy to prove that if $\varphi(\m_i) \longrightarrow_{G_{i-1}} h$ and $\varphi(t\v) \longrightarrow_{G_{i-1}} h'$, then $\lm(h) = \lm(h')$ \footnote{Assume $\lm(0) = 0$.}. Moreover, $\{h\} \cup G_{i-1}$ is a strong Gr\"obner basis of $\image_i(\varphi)$, and so is $\{h'\} \cup G_{i-1}$. Since $\varphi(t\v)$ usually has a smaller leading monomial than $\varphi(\m_i)$, reducing $\varphi(t\v)$ may cost fewer time.

\subsection{Using JPairs to avoid irreducible preimages}

Although many redundant computations are rejected by syzygy criterion and ``eventually super reducible'' criterion, there are still many redundant computations resulting from $\varphi(\m_i)$ (or $\varphi(t\v)$ from the last subsection) that is not reducible by a strong Gr\"obner basis of $\image_{i-1}(\varphi)$. Similar to Buchberger introducing critical pairs, Gao et al. use JPairs to avoid this kind of redundant computations in GVW.

For $g = \varphi(\v), g' = \varphi(\v') \in G_{i-1}$, the {\bf JPair} of $g$ and $g'$ is defined as $$t(\v, g), \mbox{ where } t\lm(g) = \lcm(\lm(g), \lm(g')) = t'\lm(g'), \mbox{ and } \lm(t\v) \succ_s \lm(t'\v').$$ $\lm(t\v)$ is called the signature of the JPair $t(\v, g)$.

In GVW \citep{Gao10b}, Gao et al. have proven that only reducing the polynomials from JPairs, is enough to build a Gr\"obner basis for $\ker(\varphi)$ as well as a strong Gr\"obner basis of $\image(\varphi)$.

After introducing JPairs, it is possible that several JPairs have the same signature. Based on the fact discussed in Subsection \ref{subsec_simpler}, the reducing results of these JPairs will have the same leading monomial. So only one of these JPairs have to be reduced in practice, and other JPairs can be rejected. The difference between F5 and GVW just lies in the strategy of rejecting redundant JPairs/critical pairs that have the same signature.

\subsection{Computing order of JPairs}

The first edition of GVW assumes JPairs are computed by an ascending order on their signatures, which is the same as the algorithm described in Subsection \ref{subsec_gvwinmmm}. The correctness of this GVW is proved in the first edition of GVW paper, and the termination is proved in \citep{Huang10}. Later, after the ``eventually super reducible'' criterion is improved, the GVW algorithm allows to compute JPairs in any order. The correctness proof of GVW in this version is given in the second edition of GVW paper, and the termination is proved in \citep{SunWang12}.

\subsection{Recovering}

For a strong Gr\"obner basis $G_{i} = \{g_1 = \varphi(\v_1), g_2 = \varphi(\v_2), \ldots, g_s = \varphi(\v_s)\}$ of $\image_i(\varphi)$, it is not necessary to store a full vector $\v$ such that $\varphi(\v) = g\in G_i$ during the practical implementation, since only $\lm(\v)$ is needed in the reductions as well as criteria. In GVW, Gao et al. give a method of recovering a full vector $\v'$ such that $\lm(\v') = \lm(\v)$ and $\varphi(\v') = g$ after the algorithm terminates. The authors modify this method to obtain Gr\"obner bases for syzygy modules directly from outputs of F5 in \citep{SunWang11c}.

\subsection{Putting all together}

Putting all the ideas discussed earlier, we get the true GVW algorithm.

\begin{algorithm}[!ht]
\DontPrintSemicolon
\SetAlgoSkip{}
\LinesNumbered

\SetKwData{jpset}{JPairSet}
\SetKwData{todo}{Todo}
\SetKwData{newltset}{NewLTSet}
\SetKwFunction{syz}{Syzygy}
\SetKwFunction{rew}{Rewritten}
\SetKwFunction{red}{Reduce}
\SetKwInOut{Input}{Input}
\SetKwInOut{Output}{Output}
\SetKwFor{For}{for}{do}{end\ for}
\SetKwIF{If}{ElseIf}{Else}{if}{then}{else\ if}{else}{end\ if}

\Input{$\varphi$, the $k[X]$-homomorphism from $k[X]^m$ to $k[X]$, defined by $\{f_1, \ldots, f_m\}$ in the Subsection \ref{subsec_frommmm}; $\prec_s$ and $\prec_p$, monomial orderings on $k[X]^m$ and $k[X]$ respectively.}

\Output{$H$, a Gr\"obner basis of $\ker(\varphi) = \syz(f_1, \ldots, f_m)$; $G$, a strong Gr\"obner basis of $\image(\varphi) = \langle f_1,\ldots, f_m\rangle$.}

\BlankLine

\Begin{

$H \lla \{f_j\e_i - f_i\e_j \mid i, j = 1, 2, \ldots, m\}$

$G \lla \{(\e_i, \varphi(\e_i) = f_i) \mid i = 1, 2, \ldots, m\}$

$\jpset \lla \{$all JPairs of $G\}$

\While{$\jpset \not= \emptyset$}
{
$t(\u, f) \lla$ a JPair in $\jpset$

$\jpset \lla \jpset \setminus \{t(\u, f)\}$

\If{there is no $\w\in H$ such that $\lm(\w) \mid \lm(t\u)$ {\bf\rm AND} $t(\u, f)$ is not rejected by ``eventually super reducible'' criterion w.r.t. $G$}
{
$t(\u, f) \longrightarrow_G (\w, h)$

\If{$h = 0$}
{$H \lla H \cup \{\w\}$}
\Else{
$\jpset \lla \jpset \cup \{$JPairs generated from $(\w, h)$ and $G\}$
$G \lla G \cup \{(\w, h)\}$
}

}
}

{\bf return} $H$ and $G$
}
\caption{The GVW Algorithm}
\end{algorithm}

\section{Conclusions} \label{sec_conclusions}

The theories of GVW as well as signature-based Gr\"obner basis algorithms are explained from the view of MMM algorithm in this paper. From this view, we try to make signature-based algorithm easier understood. 

Theories on signature-based algorithms are relatively complete now. The only problem left may be that the termination of original F5 is unproved. Besides, implementing signature-based algorithms more efficiently is also quite challenging.

\end{document}